\documentclass[prb,twocolumn]{revtex4-2}

\usepackage{}
\usepackage{epsfig}
\usepackage{graphicx}
\usepackage{amsfonts,amssymb}
\usepackage{amsmath}
\usepackage{color}
\usepackage {times}

\definecolor{refcolor}{rgb}{1.0,0.0,0.0}

\definecolor{refcolor}{rgb}{1.0,0.0,0.0}

\begin{document}

\title{Disorder driven maximum in  the magnetoresistance of spin polaron systems}

\author{Tanmoy Mondal$^1$ and Pinaki Majumdar$^2$}

\affiliation{$^1$~Harish-Chandra Research Institute 
(A CI of Homi Bhabha National Institute), 
Chhatnag Road, Jhusi, Allahabad, India 211019\\
$^2$~School of Arts and Sciences, Ahmedabad University, 
Navrangpura, Ahmedabad,
India 380009}
\date{\today}

\begin{abstract}
Ferromagnetic polarons are self trapped states of an electron in a locally
spin polarised environment. They occur close to the magnetic $T_c$ in low 
carrier density local moment magnets when the electron-spin coupling is 
comparable to the hopping scale. In non disordered systems the primary 
signatures are a modest non-monotonicity in the  temperature dependent 
resistivity $\rho(T)$, and a magnetoresistance that can be $\sim 20-30 \%$ 
at $T_c$, at fields that, in energy units, are $\sim 0.01 k_BT_c$. We find 
that structural disorder, in the form of pinning centers, promotes polaron 
formation, hugely increases the resistivity peak at $T_c$, and can enhance 
the magnetoresistance to $\sim 80\%$. The change in magnetoresistance with 
disorder is, however, non-monotonic. Too much disorder just creates an 
Anderson insulator - with the resistivity unresponsive to the magnetisation. 
This paper establishes the optimum disorder for maximising the 
magnetoresistance, suggests the physical process behind the unusual disorder 
dependence, and provides a magnetoresistance map - in terms of coupling and 
disorder - that locates some of the existing magnetic semiconductors within 
this framework.
\end{abstract}

\maketitle

\section{Introduction}

An itinerant electron coupled to the background spins in a 
local moment ferromagnet can self-trap as the magnetic 
disorder increases with increase in temperature. The  electron 
spin  acts as a magnetic field on the background moments and, under
restricted conditions, can create a polarised neighbourhood whose
`potential well' traps the electron. This self-trapped state is 
a ferromagnetic polaron (FP), the outcome of a competition between 
internal energy gain and entropy loss
\cite{ref_EuS,ref_EuS1,ref_other_Eu_materials,
ref_exp2,ref_exp4,ref_EuB6_opt_cond1}. 
The state does not form if the 
electron-spin coupling is below
a threshold, or if the electron density is not sufficiently low, 
or if the temperature is too low (in which the moments are anyway 
parallel) - or if it is too high (in which case entropy wins).
The FP is an elusive object in the parameter space.

The concept of magnetic polaron 
originated in the 1960s in the context of Eu-based magnetic
semiconductors \cite{ref_other1,ref_EuS,ref_EuS1}, and has since been
invoked to explain unusual transport behaviour in a broad class of
low–carrier-density local-moment ferromagnets. 
The key signature of FP formation is the nonmonotonic temperature
dependence of the resistivity $\rho(T)$ - 
the conventional magnetic scattering
 induced rise and saturation of $\rho(T)$ across $T_c$ is replaced
 by a peak near $T_c$, followed by an extended regime with 
 $d\rho/dT < 0$. Such nonmonotonic resistivity has been observed
 in a variety of low carrier density local moment ferromagnets
 \cite{ref_EuB6_resistivity,ref_other1,ref_other_Eu_materials,
ref_exp2,ref_exp3, ref_exp_GdSi,ref_exp4,ref_EuB6_opt_cond1}.

Most of the ferromagnetic semiconductors having polaronic features,
e.g,  EuO, GdN etc, are highly prone to defects. Defects generated
intrinsically, or introduced through doping, act as centres
for electron trapping \cite{ref_EuO_res_1,ref_EuO_res_2,ref_GdN_res,
ref_GdN_dope,ref_GdN_band1,ref_GdN_band2,ref_GdN_gap,
ref_EuO_raman,ref_EuO_meuon}. The prominent features of these 
disordered materials are, (i) a dramatic variation in the
resistivity from sample to sample, with the low
temperature resistivity
 ranging from $10^{-5}$ to $10^{-3}~\Omega$cm,
 (ii) a peak in resistivity near $T_c$ that spans several orders 
 of magnitude, from $10^{-3}$ to $10~\Omega$cm. 
This wide 
 variation is believed to be caused by changes in 
 disorder strength, impurity concentration, and 
electron density
 \cite{ref_EuO_res_1,ref_EuO_res_2,ref_EuO_map,ref_EuO_dope, 
ref_EuO_split,ref_EuO_band1,ref_EuO_band2,ref_GdN_res,
ref_GdN_dope,ref_GdN_band1,ref_GdN_band2,ref_GdN_gap}. 
The variation in the resistivity peak at $T_c$ 
correlates with the size of the  magnetoresistance (MR)  
\cite{ref_EuB6_res_field,ref_EuB6_hall,ref_EuB6_stm}.

  On the theory side there are some works that show that in the 
  clean limit above a critical electron-spin coupling ($J'_c$) and 
  below a critical electron density ($n_c$), polarons can form 
  \cite{ref_theo_other_montecarlo,ref_theory_1polaron,
ref_theo_other_variation,ref_theo_other_dmft,
ref_theory_magneto_resistance,
ref_theory_polaron_hopping,ref_EuB6,ref_clean_fp},
 and these polaronic features are enhanced in the presence 
 of impurities \cite{ref_EuO_theory_res_1,ref_EuO_theory_res_2,
ref_EuO_theory_res_3,ref_EuO_theory_res_4,
ref_EuO_theory_res_5,ref_EuO_theory_res_6,
ref_EuO_theory_res_7,ref_gdn_theory,ref_EuO_theory}.
 However, several open questions remain to be answered.
For example, (i)~are the effects of electron-spin coupling and
structural disorder additive on the resistivity (Mathiessen's
rule)?  (ii)~can disorder induce
 polaron formation when the clean problem does not
support polarons?
 (iii)~how does disorder affect the magnetoresistance? 
While it is obvious that too much disorder will make the
transport insensitive to magnetic order, and hence an
applied field, the behaviour at moderate disorder
remains unknown. Clarifying this is the central object
of this paper.

The questions above 
are difficult to address because self-trapping 
occurs at finite-temperature, in a spatially correlated spin 
background, and involves
 strong electron–spin coupling where analytical approaches 
 are inadequate. Numerical simulations, in turn, are 
 constrained by the need to treat both thermal spin 
 fluctuations and the electronic sector accurately, 
 which limit the accessible system size, and the
  inclusion of disorder adds another layer of complexity.

Avoiding too much material specificity for the moment,
in terms of band structure, etc, we study the `minimal
model' of 
a lattice of Heisenberg spins with ferromagnetic coupling $J$,
 with the spins ``Kondo coupled'' to itinerant electrons with 
 coupling $J'$ \cite{ref_theo_other_montecarlo,ref_theory_1polaron,
ref_theo_other_variation,ref_theo_other_dmft,
ref_theory_magneto_resistance,
ref_theory_polaron_hopping,ref_EuB6,ref_clean_fp,ref_EuO_theory}. 
The electrons have a tight-binding hopping $t$ and experience
 an attractive potential $-V$ at a fraction of sites. 

While the effect of the `quenched' impurity potential can be
readily handled numerically, the difficulty is in computing 
the effect of the electrons on 
 the core spins.
We use a Langevin dynamics (LD) scheme 
\cite{ref_EuB6,ref_clean_fp,ref_EuO_theory,ref_spin_lang,
ref_spin_lang1,ref_chern}  to treat thermal fluctuation of
the spins, with an exact diagonalisation method to compute
the torque.
An alternate Monte Carlo approach to these models has
 typically been limited to size  $\sim 10 \times 10$, but
 the parallelised LD approach can access lattices with
  $N \sim$ $20 \times 20$. This size, though still modest,
allows us to study a finite concentration of polarons and map
their transport, spatial, and spectral features.

In this paper, we fix the electron-spin coupling, the impurity
density, and the electron density and vary only the strength 
of the impurity potential $V$. We study the temperature
dependence of physical properties for a range of $V$ 
values, and the magnetic response near $T_c$. 
Our results are in a two dimensional system, we have 
commented on the dimension dependence elsewhere.
Our main results are the following:
\begin{enumerate}
\item 
Increase in $V$ leads to a rapid
increase in the resistivity peak at $T_c$, but also 
progressively insulating behaviour at both low and high $T$.
The fraction of resistivity
arising from the interplay of structural disorder and magnetic
scattering, however, first increases and then decreases with
$V$.
\item
The magnetoresistance at $T_c$, for a fixed field, 
initially grows  with increasing $V$, peaks at some
$V_{opt}$ and then decreases again. The maximum 
magnetoresistance, at $h \sim 0.1T_c$,  
can be $\sim 90\%$.
\item 
A pseudogap develops at weak disorder and deepens with increasing disorder, but is suppressed by a magnetic field; at stronger disorder, disorder-induced localisation dominates, leading to a field-independent gap.
\item
The resistivity can be understood from the
character of electronic states close to the chemical 
potential. At weak disorder, the electronic states at 
energy $\epsilon \lesssim \mu$ are delocalised at 
all $T$. For $V \sim V_{opt}$ the mobility edge 
crosses $\mu$ as $T \rightarrow T_c$. This `insulating'
system can be metallised by an applied field. At large 
$V$ the mobility edge is well above $\mu$ at all $T$ 
and a field cannot cause a transition. The large MR 
is a field driven insulator-metal transition.
\item
We constructed a `magnetoresistance map' in terms of coupling $J'$ and
potential $V$ to delineate the transport regimes and tentatively
locate experimental samples within this scheme. We find, remarkably,
that $V_{opt} \approx 2t$ and is only weakly dependent on $J'$.
\end{enumerate}

\section{Model and method}

We study the Heisenberg-Kondo (H-K) model in 2D with 
disorder as impurity sites as:
\begin{equation}
H = 
 -t \sum_{\langle ij \rangle ,\sigma} 
c^{\dagger}_{i\sigma} c_{j\sigma}
- J'\sum_i {\bf S}_i.{\vec \sigma}_i 
-J \sum_{\langle ij \rangle} {\bf S}_i.{\bf S}_j + \sum_i V_i. n_i
\end{equation}
We set the potential $V_i = V$ on a fraction $n_{imp}$ of
 randomly chosen lattice sites and zero elsewhere. We set 
 $t=1$, $ J=0.01t$, electron density $n_{el}=0.025$ and
 $n_{imp}=0.02$. We varied
  $V$ in the range $0.5t \leq V \leq 4.0t$.
To generate equilibrium spin configurations
$\{{\bf S}_i \}$ we use a LE where
the spins experience a 
torque (below) derived from $H$ alongside 
a stochastic kick, with variance $\propto k_BT$, 
to model the effect of temperature.  
The LE has the form:
\begin{eqnarray}
\frac{d\mathbf{S}_i}{dt} &=& \mathbf{S}_i \times (\mathbf{T}_i 
+ \mathbf{h}_i) - \gamma \mathbf{S}_i \times (\mathbf{S}_i 
\times \mathbf{T}_i) \cr
\mathbf{T}_i &=& -\frac{\partial H}{\partial \mathbf{S}_i} =
-J' \langle \vec{\sigma}_i \rangle -
J \sum_{\delta} {\bf S}_j \delta_{j,i+\delta}
\cr
\langle {h}_{i \alpha} \rangle &=& 0,~~ 
\langle
 h_{i \alpha}(t)h_{j \beta}(t') \rangle = 2\gamma k_B T
 \delta_{ij} \delta_{\alpha \beta} \delta(t-t') 
\end{eqnarray}
 $\mathbf{T}_i$ 
is the effective torque acting on the spin at the $i$-th 
 site, $\gamma =0.1 $
 is a damping constant, and $\mathbf{h}_i$ 
is the thermal noise satisfying the fluctuation-dissipation
 theorem.
$\langle \vec{\sigma}_i \rangle$ represents the 
expectation of $\vec{\sigma}_i$ taken over the
instantaneous ground state of the electrons, and
$\delta$ is the set of nearest neighbours. 
In the presence of an external magnetic field there is
an additional term ${\hat z} h$ in ${\bf T}_i$. 
Once Langevin evolution reaches equilibrium the magnetic
correlations can be computed from the 
$\{{\bf S}_i \}$, and electronic features obtained
by diagonalisation in these backgrounds.

\section{Results at zero field}

Fig.1(a) shows $\rho(T)$ for different $V$.
For $V=t$ the nonmonotonic behaviour of near $T_c$ is enhanced 
 compared to a clean system \cite{ref_clean_fp,ref_EuB6}.
 A disorder-driven insulating tendency 
 appears as $T \rightarrow 0$, but this feature is not 
 related to polaronic physics.
For $V=2t$, the nonmonotonicity near $T_c$ is further 
 amplified.  The combined 
 influence of $V$ and $J'$ is not a simple addition 
 of their individual effects; rather, the system responds 
 in a strongly intertwined and nonlinear manner.
 Moreover, a pronounced insulating behaviour 
 emerges as $T \rightarrow 0$, with a much stronger 
 insulating character than in the $V = t$ case.
If the disorder strength is increased further, the electrons
 become strongly trapped around the impurity sites,
 and insensitive to the magnetic background. 
 As a result, the polaronic effect is progressively 
diminished.  The strong localisation around impurities 
leads to an insulating $\rho(T)$ behaviour at all $T$,
with  $\rho(T)$   determined mainly by $V$. 

% -----------------------------------------------------
\begin{figure}[b]
\centerline{
\includegraphics[height=5.2cm,width=8.5cm]{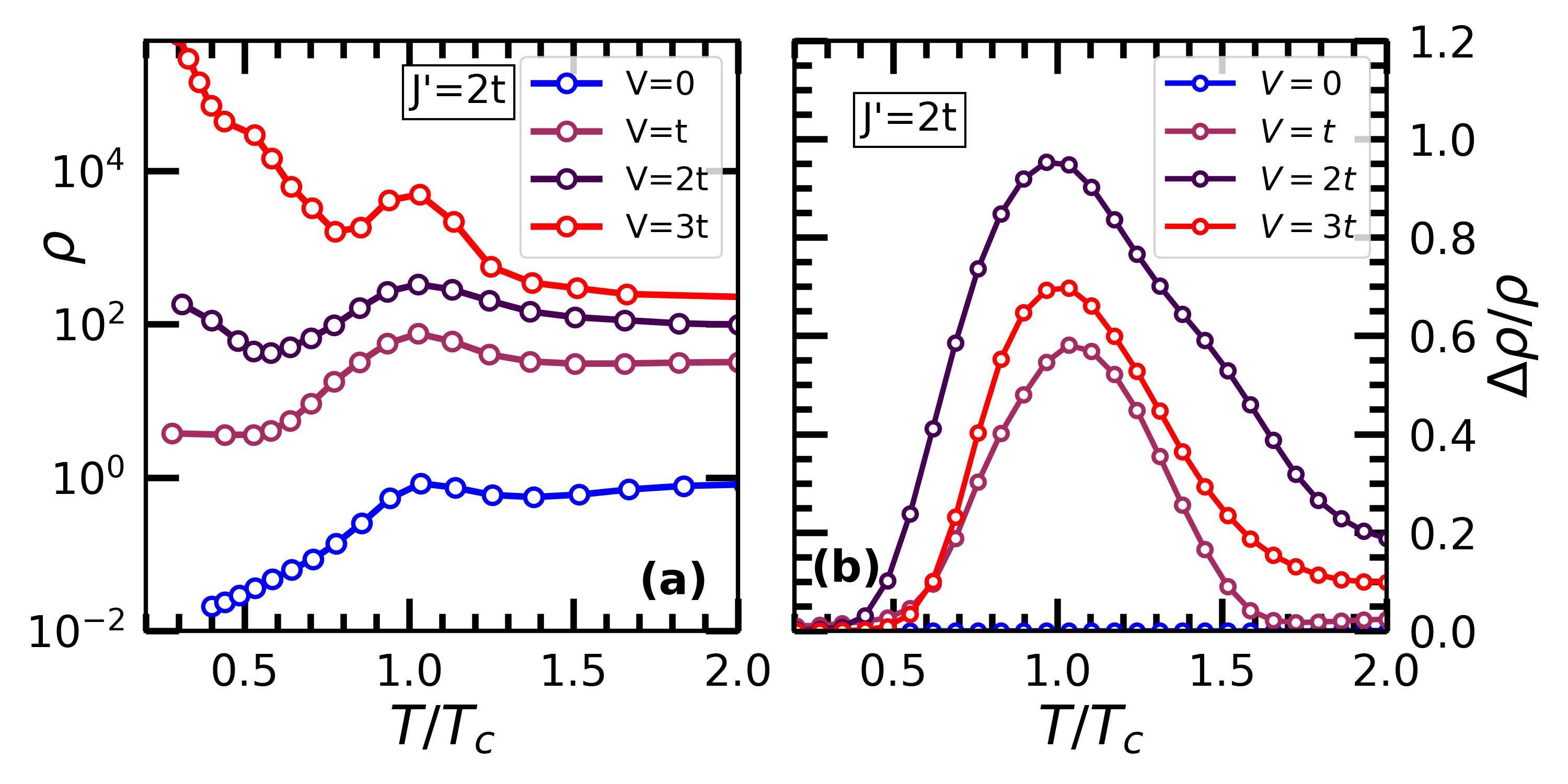} }
\caption{The zero field resistivity $\rho(T)$. (a)~The 
resistivity at $V=0$ and three finite values of $V$. 
Note the maxima-minima structure in $\rho(T)$ and the 
increasing insulating behavior at low and high temperature
as $V$ increases. (b)~The `cross term' $\rho_{cr}(T)$ (see
text) normalised by $\rho(T)$. This captures the 
enhancement due to the coupling of structural and
magnetic effects. This function grows initially with 
increasing $V$, and drops after $V = 2t$.}
\end{figure}
% -----------------------------------------------------
\begin{figure}[t]
\centerline{
\includegraphics[height=7.5cm,width=8.5cm]{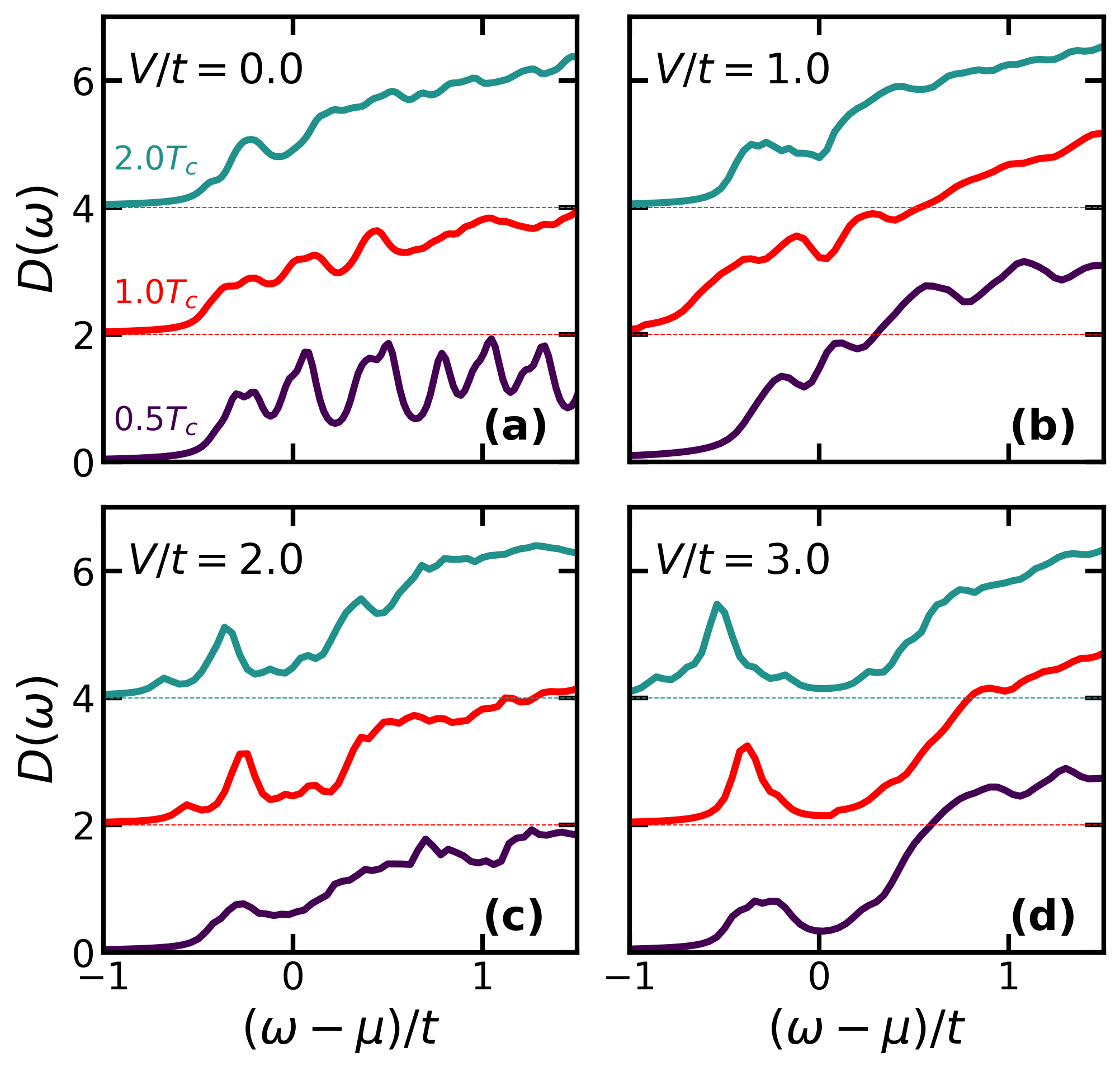} }
\caption{Density of states at zero field. Panels (a)-(d)
show the DOS for $V/t=0,1,2,3$, respectively, for
three values of $T/T_c$ (marked in panel (a)). 
Other panels follow the same colour code. At $V=0$ 
there is essentially featureless at all $T$, with the
finite size fluctuations smoothing out with $T$. At 
$V/t=1,2$ we see a depression (pseudogap) appearing
around $\omega \sim \mu$ with increasing $T$. At $V=3t$
there is a prominent depression already at $T=0$ and
it becomes somewhat broader at persists to the highest
$T$. We are approaching a regime where the DOS is
`gapped' and essentially $T$ independent.  }
\end{figure}
% -----------------------------------------------------

We define the resistivity $\rho_{cr}(T)$ that arises from
the cross coupling of impurity and magnetic effects as 
below:
$$
\rho(T) = \rho_V(T) + \rho_{J'}(T) + \rho_{cr}(T)
$$
where $\rho(T)$ is the full resistivity, $\rho_V(T)$ is the
resistivity with $V$ acting alone, and $\rho_{J'}(T)$ is
the resistivity of the clean magnetic system. $\rho_{cr}(T)$
then encodes the enhancement of resistivity due to the
`interference' of structural and magnetic effects. And,
it is this `extra' that can be affected by a magnetic
field to generate a MR that is bigger than in the clean
system. Fig.1(b) shows the behaviour of $\rho_{cr}(T)/\rho(T)$
for three values of $V$. By definition, this object is zero
at $V=0$.

% --------------------------------------

 We now examine how this enhanced polaron formation
 influences the DOS. Fig.2 shows the
 effect of varying $V$ and $T$ at $J'=2t$  on the DOS.
 Panel (a) shows the non disordered case for reference.
  At  $V=t$ a faint pseudogap appears at
 the Fermi energy. This reflects
 the strengthening of electron localisation and the
 formation of stronger FPs, which generate bound
 states near the lower edge of the DOS. 
At $V=2t$, the pseudogap deepens and becomes more
 pronounced.
  By $V=3t$, a clear gap emerges.
 In this strong-disorder limit, the FP picture
 becomes less relevant: electron motion is
 dominated by disorder-induced localisation,
  which governs the system’s behaviour more
  strongly than polaronic effects.

\section{Effect of a magnetic field}

Upto now, we have talked about the zero field resistivity. Now we
discuss the effect of applied field. Application of a magnetic field
aligns the localised spins toward ferromagnetic order, thereby
suppressing the nonmonotonic behaviour of $\rho(T)$. For
sufficiently large fields, the background spins become fully
ferromagnetically aligned, and the nonmonotonicity is
completely removed and which gives rise to colossal MR.
So, the stronger the nonmonotonicity the larger the MR. The MR can be quantified through the ratio $(\rho(T_c,h=0)-\rho(T_c,h))/\rho(T_c,h=0)$. However, a simplified estimate of MR can be obtained by replacing $h$ with $h_c$, where $h_c$ denotes the field strength required to eliminate the nonmonotonicity. In practice, this can be estimated more conveniently by
assuming a fully ferromagnetic spin background,
effectively replacing $h_c$ with $h_{\inf}$.

Fig.3(a) shows the effect of $V$ and $h$ on $\Delta \rho$ at $J'=2t$. For a fixed $h$, $\Delta \rho$ initially increases with disorder strength $V$, reaches a maximum at a characteristic value $V_{opt}$, and
subsequently decreases. This behaviour arises because,
at low and intermediate disorder, the nonmonotonicity of
$\rho(T)$ strengthens with increasing $V$, enhancing the polaronic contribution to MR. Beyond $V_{opt}$, however, strong disorder progressively lessens the influence of the magnetic background, suppressing polaronic features and thereby reducing MR. 
As expected, the magnitude of the MR depends strongly on $h$, showing a $75\%$ increase as the field is increased from $0.001t$ to $0.01t$.

Figure 3(b) shows the $J'$ dependence of MR estimated within the high-field calculation. The MR increases gradually with increasing $J'$, as expected from the corresponding enhancement of the nonmonotonicity with $J'$. Although the rate of increase slows down at larger $J'$. The location of $V_{opt}$ exhibits only a weak dependence on $h$.

% -----------------------------------------------------
\begin{figure}[b]
\centerline{
\includegraphics[height=4.9cm,width=8.5cm]{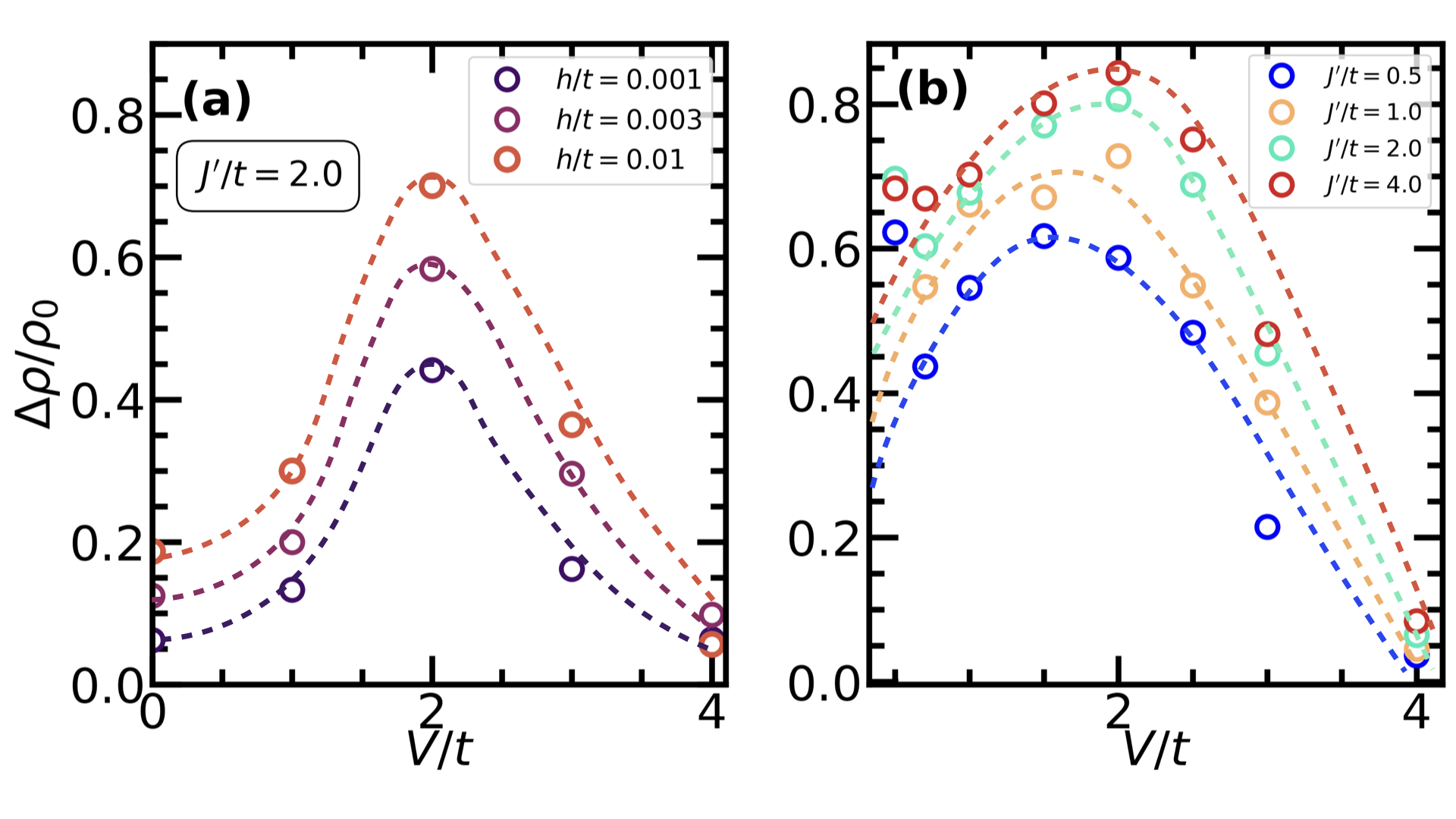}
}
\caption{Magnetoresistance. (a)~The MR, defined as 
$
(\rho(T,0) - \rho(T,h))/\rho(T,0)$ at $T=T_c$ and 
$h/t = 0.001,0.003,0.01$ shown for varying $V$.
The MR expectedly increases with increasing $h$, but has an unusual $V$
dependence. The moderate MR, $\sim 10-20\%$ at $V=0$ increases to a maximum
$\sim 60-70\%$ at $V \sim 2t$, and then falls off.
(b)~The high temperature, high field, MR - $(\Delta \rho/\rho)_{\infty}$
- where $\rho$ is the resistivity is the background of the structural
disorder and random spins, and the suppression corresponds to 
a fully spin-polarised background. This calculation does not
involve any magnetic annealing. We want to highlight that
this quantity is a reasonable proxy for the finite $T$,
finite $h$ MR both in terms of magnitude and $V$ dependence.}
\end{figure}
% -----------------------------------------------------

% -----------------------------------------------------
\begin{figure}[t]
\centerline{
\includegraphics[height=7.5cm,width=8.5cm]{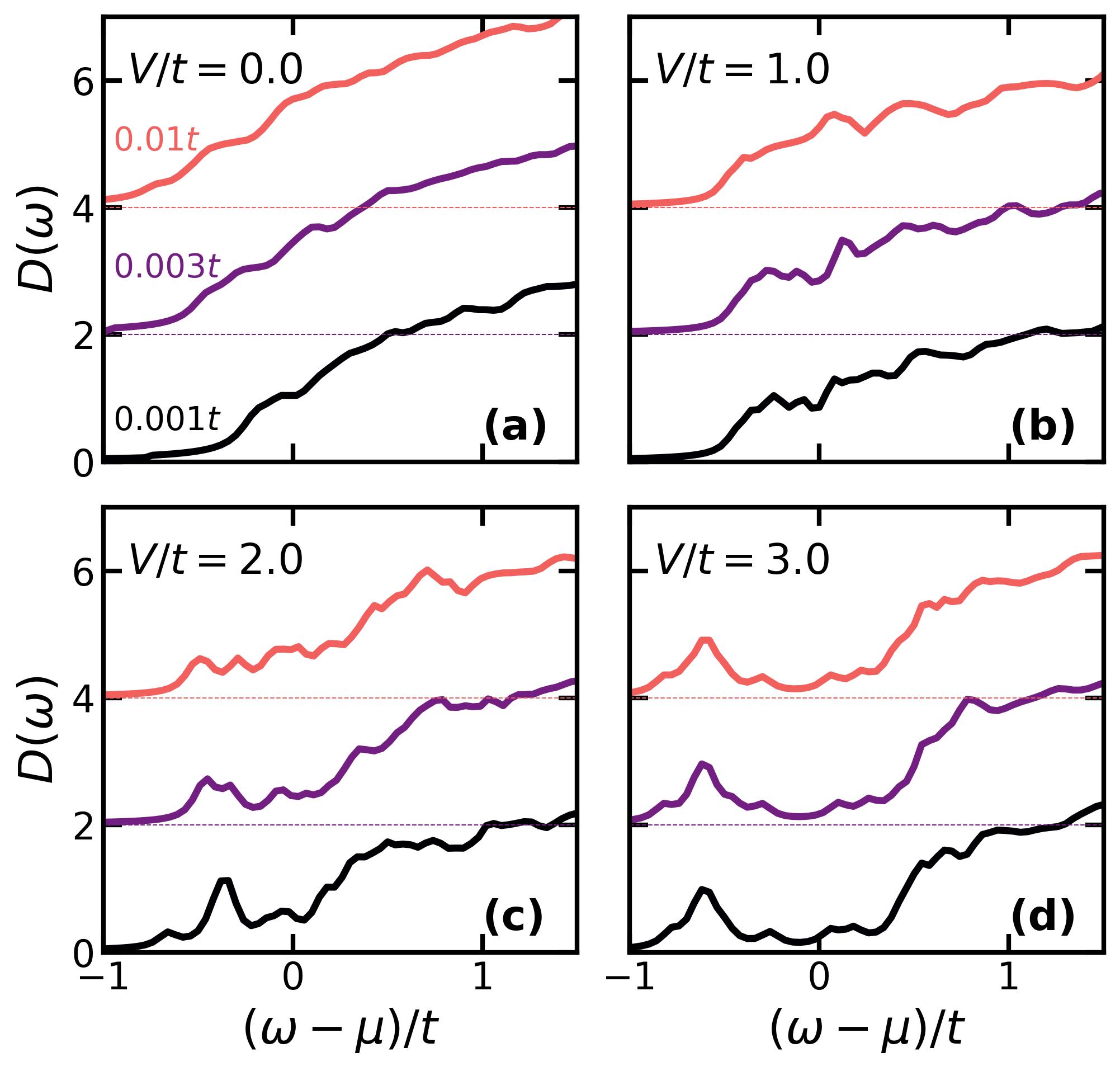} }
\caption{Field dependence of the density of states. The panels
(a)-(d) show the DOS at $T=T_c$ for $h/t=0.001,0.003,0.01$ and
$V/t=0,1,2,3$. (a)~At $V=0$ the DOS is essentially featureless
and insensitive to $h$. (b)-(c) The modest dip at $\omega \sim \mu$
at $h=0$ is smoothed out with increasing $h$. For $V=2t$ the
low $h$ dip is more prominent than at $V=t$. (d)~The prominent
suppression in the DOS near $\mu$ at $T_c$ is hardly affected
by application of the field. }
\label{fig_dos_h}
\end{figure}
% -----------------------------------------------------

We now discuss the effect of a magnetic field on the DOS for different values of $V$ at $T=T_c$, as shown in Fig. \ref{fig_dos_h}. In the absence of disorder [panel (a)], no pseudogap-like feature is present, and consequently, the effect of the magnetic field is unobservable. In contrast, for stronger disorder, where a pseudogap develops at the Fermi level at $h=0$, the application of a magnetic field suppresses the gap, as seen for $V=t$ and $2t$. This behaviour indicates the suppression of polaronic effects with increasing $h$. However, for $V=3t$, the gap remains essentially unaffected by the field. In this strong-disorder limit, the influence of localised spins becomes much less significant, rendering the FP picture less relevant.

\section{Insulator-metal transition scenario}

We now try to identify the basic mechanism behind
the transport and spectral response of the system and
frame it in a familiar language.
There are no electron-electron interactions in our
model and the physics can be understood in terms of
the single particle eigenstates, albeit in complex,
annealed, background configurations $\{ V_i, {\bf S}_i \}$.
While the $V_i$ are specified, the $\{{\bf S}_i \}$ 
distribution depends on the $V_i$ as also $T$ and $h$.

As the spin distribution evolves from its $T=0$ full polarisation, 
through the $T \sim T_c$ locally polarised configurations,
to the random $T \gg T_c$ state, the mobility edge shifts
with $T$. At $T=0$ the electrons are subject only to structural
disorder, and ideally in the 2D geometry where we are studying
the problem all states in the band would be localised. We are
however considering only states with localisation length $\ll L$,
to mimic what one would see in 3D. With this caveat, the mobility
edge $\epsilon_c$ starts below $\mu$ at small $V$ and low $T$
and moves up as $T$ increases. 
At weak $V$ it does not cross $\mu$ at any $T$, approaching it
closest near $T_c$, Fig.5(a). 

% -----------------------------------------------------
\begin{figure}[b]
\centerline{
\includegraphics[height=5.5cm,width=8.0cm]{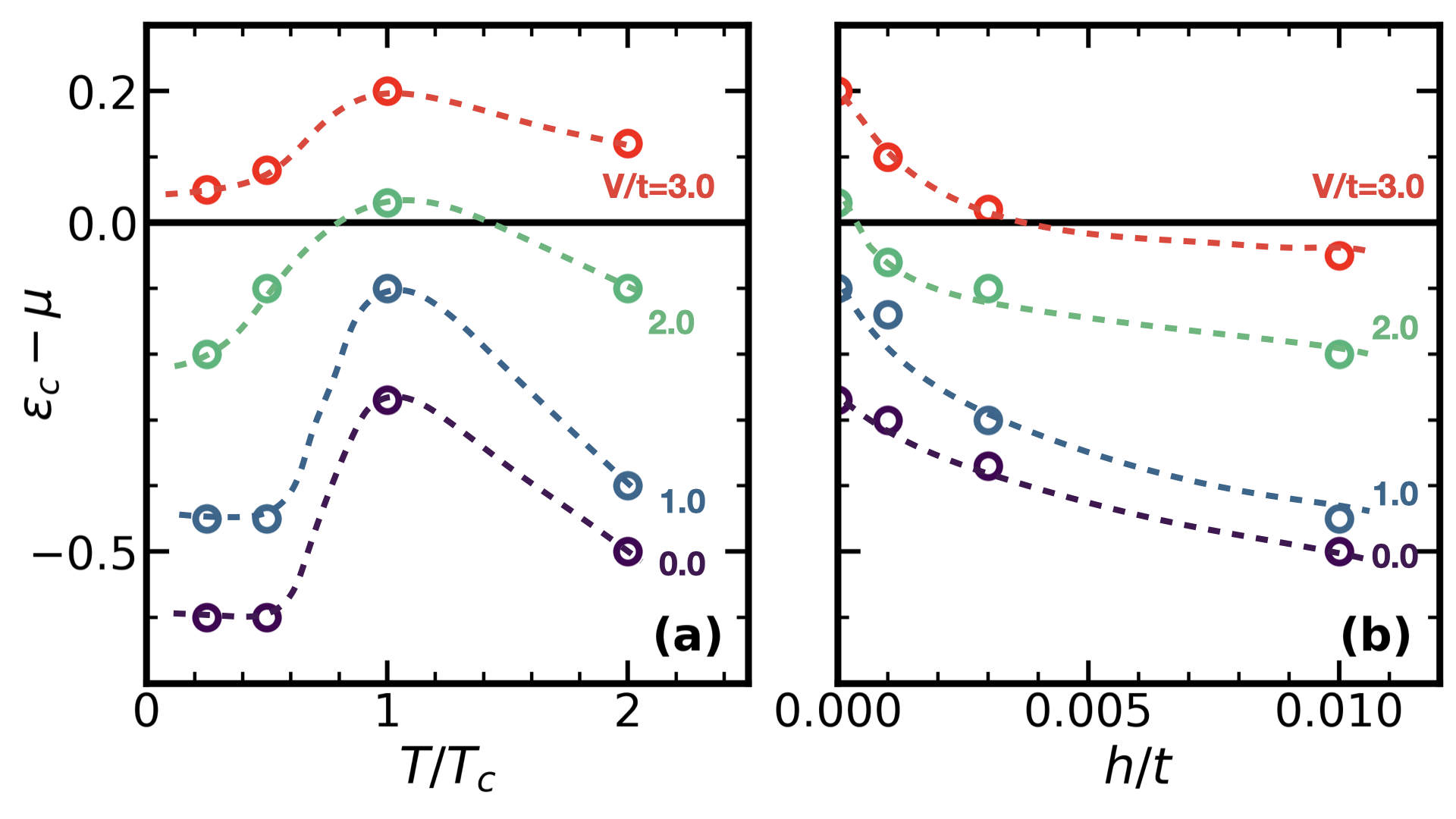} }
\caption{Location of the mobility edge $\epsilon_c$ with
respect to the chemical potential. The mobility edge is
calculated via an estimate of the inverse participation ratio
(IPR) of the electronic states on the annealed backgrounds. 
(a)~Shows
the temperature dependence of $\epsilon_c - \mu$ 
at $h=0$ for $V/t = 0,1,2,3$. For $V/t=0,1$ the mobility
edge approaches $\mu$ from below, comes closest for
$T \sim T_c$, but never crosses. The states are always
delocalised. At $V=2t$ it crosses near $T = T_c$ and 
drops back again. For $V=3t$ $\epsilon_c - \mu > 0$ for
all $T$, the states at the chemical potential are always 
localised.  
(b)~This tracks the field dependence of $\epsilon_c - \mu$ 
staying at $T=T_c$. The applied field expectedly drives the
mobility edge downwards since it weakens localisation. From
the data we see that at $V=2t$ a weak field can push $\epsilon_c$
from above $\mu$ to below $\mu$, crudely causing an `insulator-metal
transition'. Something similar happens at $V=3t$ as well, but at
much larger field. At even larger $V$ the insulating state will
persist at arbitrarily large $h$.  }
\end{figure}
% -----------------------------------------------------
% -----------------------------------------------------
\begin{figure}[t]
\centerline{
\includegraphics[height=5.5cm,width=7.2cm]{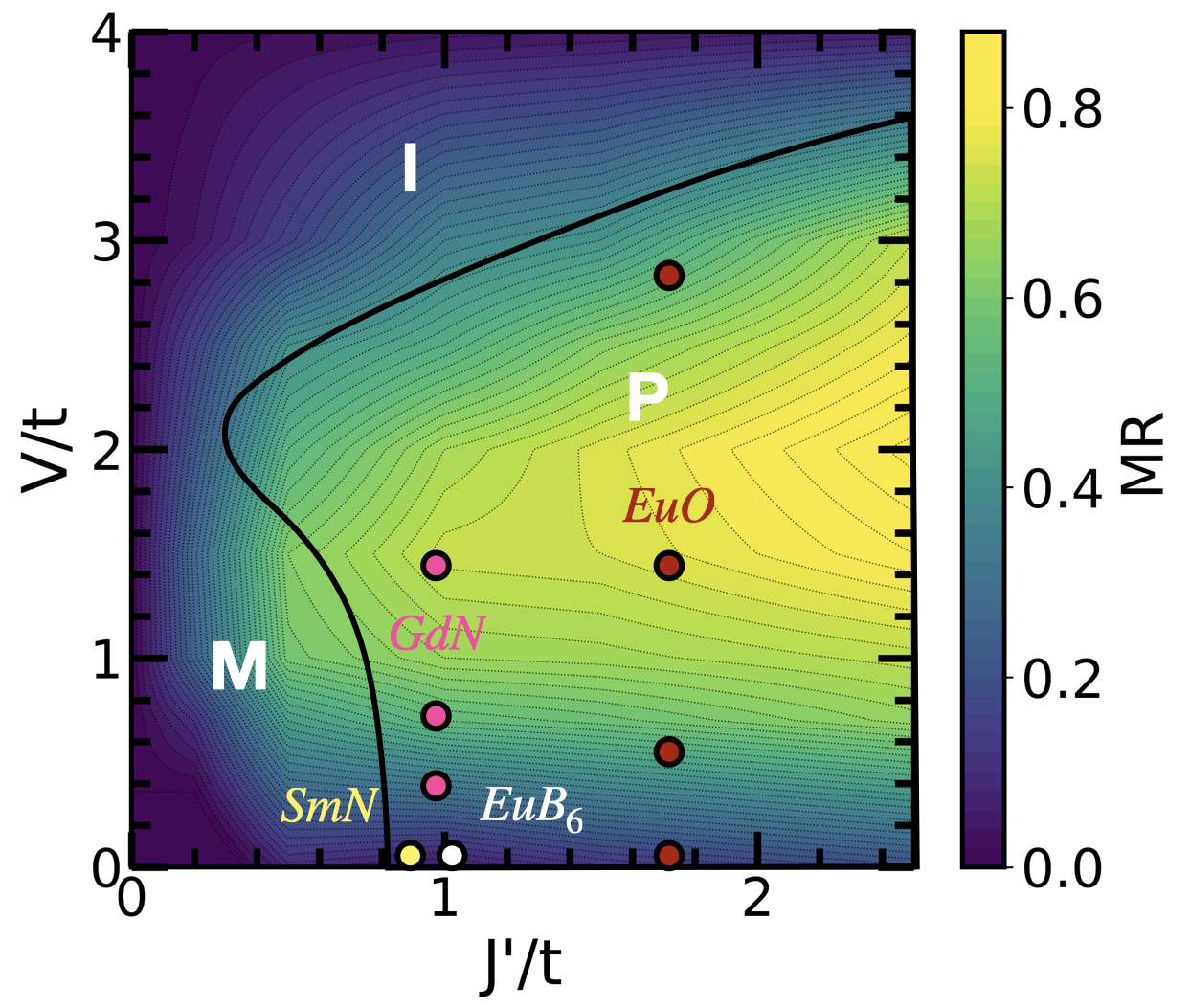} }
\caption{Map of magnetoresistance. This panel shows
a surface plot of the `magnetoresistance' 
$(\Delta \rho/rho)_{\infty}$, which, we have demonstrated,
is a good mimic of the fully annealed result. At a typical
$J'$ increasing $V$ leads to an increase in MR and then
a fall. The optimal disorder is $V \sim 2t$, slowly increasing
with $J'$. The firm black line separates the polaronic regime (P),
where we have a peak-dip structure in $\rho(T)$, from the
simple metallic (M) or insulating (I) regimes.}
\end{figure}
% -----------------------------------------------------

We see enhanced scattering, and a peak in resistivity, but
no `localisation' of states at $\mu$ even at $T_c$ for $V=0,t$.
At $V=2t$ however the interplay of $V_i$ and polaron formation
leads to $\epsilon_c$ crossing $\mu$ near $T=T_c$. Nominally
we would have a mobility gap but the finite $T$ leads to a
finite conductivity. At $V=3t$, however, $\epsilon_c$ is
above $\mu$ at all $T$ and the chemical potential is
in the region of localised states. We have an insulator
with $t$ dependent mobility gap. On the whole, the
scenario we propose (modulo 2D weak localisation effects)
is that both structural disorder and thermally induced
polaron formation can drive a metal-insulator transition in the
system.

Fig.5(b), where we consider the field dependence of $\epsilon_c
- \mu$i at $T=T_c$, is just the reverse. Now at $V=0, t$, the
mobility edge is pushed below, towards the lower band edge as
$h$ increases, making the states near $\mu$ more conducting.
At $V=2t$, where the $h=0$ mobility edge was above $\mu$ we 
see that a small field can push $\epsilon_c$ below $\mu$.
This would now be a field driven insulator-metal transition,
a complement of the thermally driven metal-insulator transition.
We see echoes of manganite physics but with very different
microscopic ingredients. At $V=3t$ there is an effect similar
to what we see at $V=2t$, but now at a much higher field.
At even larger $V$ we surmise that even very large $h$ would
not be able to change the insulating character.

This analysis provides a pointer to the $V_{opt}$ that maximises
the MR. We need a $V$ such that at $h=0$ the mobility edge just
crosses $\mu$ for $T \sim T_c$. In that case a weak field will
be able to push it below $\mu$, create an insulator-metal
transition, and generate large MR. At $V < V_{opt}$ the mobility
edge never crosses $\mu$, for $V \gg V_{opt}$ the mobility edge
cannot be pushed down below $\mu$.

\section{Phase Diagram} 

While our detailed discussion has been at a `typical'
electron-spin coupling $J'=2t$, with varying $V$,
we would like to put forward a broader phase diagram
based on an approximate calculation. Gig.6 attempts
this, where, for the parameters used in this paper,
but varying $J'$ and $V$, we suggest a magnetoresistance
map, as well as a transport regime classification.

The resistive character of the $V-J'$ problem 
can be classified into the following categories, based on
their trends at $T \ll T_c$, around $T_c$, and at $T \gg T_c$.
First, the two simple regimes are:
{\bf 1.}~Metallic (M), i.e, $d \rho/dT > 0$  at all $T$:
For $V \ll t$ and $J' < J'_c$, $\rho(T)$ is completely monotonic,
 with $\rho \rightarrow 0$ as $T \rightarrow 0$. This
 regime is well established in
 the literature on ferromagnetic materials \cite{ref_theo_res_fisher}.
{\bf 2.}~Insulating (I), i.e, $d \rho/dT < 0$  at all $T$:
At very large $V$ the magnetic background plays
 no significant role. Electrons become strongly localised around
 impurity sites, and $\rho(T)$ is insulating at all temperatures.
 The nonmonotonic feature disappears entirely, and the resistivity
 becomes essentially independent of $J'$.
The M and I regimes are marked in the figure. We have not
shown an Anderson transition point at $J'=0$ since that
is not meaningful in a 2D problem. There would be such a point
however in the 3D case, separating the M and I.

The complicated regime is {\bf 3.}~Polaronic near $T_c$:
For $V \ll t$ the resistivity is metallic both
for $T \ll T_c$ and $T \gg T_c$, while a clear $d\rho/dT < 0$
 region appears near $T_c$, signaling the formation of
 polarons.  This resembles the behaviour seen in
 EuB$_6$, SmN, etc.  For moderate disorder ($2t \leq V < 4t$)
  the resistivity becomes strongly nonmonotonic.
  At $T \rightarrow 0$, it shows insulating behaviour due to
  impurity-driven localisation, followed by an enhanced
  polaronic feature near $T_c$, and finally a metallic
  saturation at $T \gg T_c$. Several samples of EuO,
  GdN, EuS and related compounds fall in this category
  \cite{ref_EuS1,ref_EuO_res_1,ref_EuO_res_2,ref_GdN_res}.
Upon further increasing $V$, the magnetic background
 becomes irrelevant in electronic transport.
On the whole the region enclosed by the firm black line
is the polaronic region where large MR is expected.

The colour code in the map corresponds to the artificial
`infinite temperature, infinite field' magnetoresistance
that we showed in Fig.3(b), since a full annealing based
calculation over $V$ and $J'$ is very expensive. The 
map reveals that the optimal $V$ is $\sim 2t$, for the
impurity and electron density used, and increases only
slightly with $J'$.

\section{Conclusion}

In summary, we have demonstrated that moderate 
disorder promotes magnetic polaron formation and 
enhances the magnetoresistance - which can reach upto $90\%$.
However, beyond an optimal disorder the MR starts decreasing
and vanishes as $V \rightarrow \infty$.
We trace the origin of this effect to a thermally driven
metal-insulator transition that arises from polaron
formation in a structurally disordered background, and
the field driven metallisation of this state driving the
large magnetoresistance.
We provide a map of the magnetoresistance in terms of
electron-spin coupling and impurity potential and
locate the optimal disorder for maximising the
magnetoresistance.

\end{document}